\documentstyle[version2,aps]{revtex}

\begin{document}
\draft
\centerline{\bf Quantum Chemistry, Anomalous Dimensions, and the Breakdown}
\centerline{\bf of Fermi Liquid Theory in Strongly Correlated Systems}
\bigskip
\centerline{\bf Gabriel Kotliar and Qimiao Si}
\centerline{Serin Physics Laboratory, Rutgers University}
\centerline{Piscataway NJ 08854 USA}
\baselineskip=18 true bp
\bigskip

\begin{abstract}
We formulate a local picture of strongly correlated systems as a
Feynman sum over atomic configurations. The hopping amplitudes between
these atomic configurations are identified as the renormalization group
charges, which describe the local physics at different energy scales.
For a  metallic system  away from half-filling,
the fixed point local  Hamiltonian is  a generalized Anderson impurity
model in the mixed valence regime. There are
three types of fixed points:
a coherent Fermi liquid (FL) and two classes of self-similar
(scale invariant)  phases
which we denote  incoherent metallic states (IMS).
When the transitions between the atomic configurations proceed
coherently at low energies,  the system is a Fermi liquid.
Incoherent transitions between the low energy atomic configurations
characterize the incoherent metallic states. The initial conditions
for the renormalization group flow are determined by
the physics at rather high energy scales. This is the domain of local
quantum chemistry. We use simple quantum chemistry estimates to
specify the basin of attraction of the IMS fixed points.

\end{abstract}

\bigskip
{\it To appear in the Proceedings of the European Physical Society
Meeting, Regensburg, Germany, March 1993}

\bigskip
\subsection{Introduction}

The anomalous normal state properties in the high ${\rm T_c}$
superconductors have led to current theoretical interests in
addressing how electron-electron interactions can result in non-Fermi
liquid fixed points.\cite{pwa,vr} For interacting fermion systems in
one dimension, the breakdown of Fermi liquid theory occurs for
infinitesimal values of the interaction and can therefore be studied
using a renormalization group (RG) which is {\it perturbative} in the
interaction strength.\cite{lutt}

In more than one dimension, such a perturbative RG analysis has led to
the conclusion that, Fermi liquid theory  describes weakly interacting
systems with regular dispersion.\cite{shankar} Therefore, the
breakdown  of Fermi liquid theory,  if it occurs, is necessarily
non-perturbative in
the interaction strength. Most of the previous approaches to this
problem, however, start with quasiparticle states and ask how
singularities can occur in the interaction vertex between these
quasiparticles. Motivated both by the non-perturbative character of
the problem  and by the  {\it local} character of the interactions we
approach the problem of the breakdown of Fermi Liquid theory  from the
opposite end: the expansion around the atomic limit suggested in
Hubbard's early work.\cite{hubbard} The main difficulties with this
approach is that a direct perturbation expansion around the atomic
limit is a classic example of a singular perturbation problem which
requires a zeroth order solution before the perturbation expansion can
be attempted. This zeroth order solution is not available except at
very high temperatures. For these reasons, this approach was largely
abandoned by most of the condensed matter community. However,
recent non
perturbative solutions of lattice models in large dimensions have
revealed that in certain regime of parameters the Hubbard
description becomes asymptotically valid at low
energies.\cite{mott,sk}
An important technical advance was the extension of early work
of Haldane on the Anderson model,\cite{haldane} through which we have
developed a systematic renormalization group method for arbitrary set
of local configurations. This allows us to control the perturbation
expansion around the atomic limit and, in the regime where non-trivial
phases occur, calculate with some rigor the low energy properties of
incoherent metallic states (IMS).\cite{sk}
We stress that at the end one ought to be able to express our results,
derived using a local method, in a language that would make contact
with perturbative calculations in the interactions. When this is done
we may be able to identify some similarities  between our ideas and
those of the theories cited earlier. In particular, we stress that the
phases we identify exhibit charge spin separation and power law type
correlation functions. These properties have striking analogies to the
Luttinger liquid in one dimension,\cite{lutt} and to the uniform RVB
in the gauge theory formulation.\cite{zou}

In this paper we outline a general framework for
determining the conditions of the breakdown of Fermi liquid theory,
{\it from a local analysis}. We explain the physical meaning of
the parameters which control the local aspects of this phenomena,
i.e. the RG charges, as well as the reasons why they are scale
dependent. In an incoherent metallic state
the evolution of the RG charges as a function of length scale, can be
determined  analytically {\it at sufficiently low energies}. At high
energies the initial conditions of the flows can be determined by
simple perturbative quantum chemistry calculations. Assuming that the
two regimes can be matched smoothly we specify the explicit parameter
range for which a two band model falls in  the basin of attraction of
the incoherent metallic phase.

\subsection{The Local View of a Metallic System}

The {\it local} description of a strongly correlated metal starts from
selecting   a relevant set of ``atomic configurations'' in a cluster
of atoms. The simplest realization of this idea would take a cluster
consisting of a single site with a small set of atomic configurations.
However, more general choices (larger clusters or more general
molecular configurations) are possible. Locally the dynamics within
this Hilbert space is described by  carrying out a Feynman sum over
all the local histories. Each history   is described by a succession
of hops between the atomic configurations. The weight of each path
depends on the local physics and the coupling to the neighboring
sites.

This idea can be made precise, in any number of dimensions, by
following Georges and Kotliar\cite{gk} in writing down a path integral
representation of the partition function and integrating out all the
electrons except for those in the selected cluster.
The result of this process is an effective action which describes the
influence of the environment on the quantum tunneling between the
configurations (i.e. the ``atomic states'') of the cluster.
Formally,\cite{gk}

\begin{eqnarray}
{\rm e}^{-{\rm S_{eff}}( \Gamma )} = \int {\Pi}' d \psi^+ d\psi
{\rm e}^{-{\rm S} [ \psi^+ , \psi ] }
\label{intout}
\end{eqnarray}
where ${\Pi}'$ involves all sites except the one we are interested,
and

\begin{eqnarray}
{\rm S_{eff}}( \Gamma ) = \sum_{n=2}^{\infty} \int d\tau_i ds_j
\Gamma_{\alpha_1 ...
\alpha_n \beta_1 ...\beta_n}^{\tau_1 ...\tau_n s_1 ...s_n}
\psi_{\alpha_1}^+ (\tau_1) ... \psi_{\alpha_n}^+ (\tau_n)
\psi_{\beta_1} (s_1) ... \psi_{\beta_n} (s_n)
\label{effact}
\end{eqnarray}
where $\psi_{\alpha}$ are orbitals of interest.

The exact determination of the various $\Gamma_{\alpha}$ for models in
finite spatial dimensions is a difficult problem. In infinite
dimensions,\cite{gk} the effective action simplifies and contains only
a quadratic term, and  quartic terms which  are given by the bare
local interaction. In the absence of broken spatial symmetries,
the coefficient of the quadratic term, which in previous publications
was denoted by $-{G_0}^{-1}$, can be simply related to the local
correlators calculated with this action. This paves the way for exact
solutions of correlated models in large dimensionality. We stress,
however, that at this point our general framework is not
restricted to large dimensions. What the large dimensionality limit
allows us to do is to calculate {\it explicitly} the effective actions
and the local correlation functions. In addition, from the knowledge
of the local physics one can reconstruct all the lattice correlation
functions.\cite{vollhardt,gk}

This local view opens the road to a classification of all the possible
low energy behaviors of metallic systems in the limit of infinite
dimensions. In finite dimensions this local analysis can give a
qualitative understanding of the nature of the metallic state. The key
point is that, while generically the problem associated with $S_{eff}$
involves many  configurations, and hence is not universal and belongs
to the realm of quantum chemistry, at low
energies the physical content of $S_{eff}$ simplifies since it has to
describe the fluctuations between a very small number of low energy
configurations. If the configurations have the same charge, we are in
an insulating situation. The typical example for such a case is
provided by the Mott insulating phase which has been studied recently
in infinite dimensions.\cite{mott} If the system is metallic, it has
to fluctuate between low energy configurations having different
charges. A natural possibility is that, the low energy configurations
with different charges are a spin singlet and a spin doublet. If the
$S_{eff}$ describes a system interacting with a bath with a non singular
density of states, one can write down the most general model which
contains a singlet and a doublet interacting and hybridizing with a
bath: this is the generalized Anderson impurity model in the mixed
valence regime.

{}From the local point of view, the survival or the demise of Fermi liquid
theory depends on how the quantum mechanical hopping amplitudes
are renormalized as we go to lower energies.
If these quantum mechanical hopping amplitudes are renormalized to
infinity, coherent hopping results in at long time scales, and the system
is a Fermi liquid. If they are
renormalized to zero, at long time scales the transitions between
the local states occur incoherently, and the resulting state is an
incoherent metallic state (IMS). That a transition between incoherent
and coherent tunneling does occur in two level systems coupled to an
environment, is well understood in the context of the macroscopic
quantum tunneling.\cite{leggett} To extend this idea to the strong
correlation problem, we identify the coordinate that performs the
quantum tunneling with the atomic configurations of the
correlated system. We then carry out an RG analysis describing the
renormalization of the tunneling amplitudes as a function of length
scale.

The minimal model in this case is a three level system, with
a singlet state and a spin doublet. The fact that the coordinate lies
in a three dimensional space brings together the physics of spin and
charge fluctuations, and gives rise to a novel zero temperature
quantum phase transition, the mixed valence critical point, which is
in a new universality class \cite{sk}. A detailed analysis of this critical
point can be found in our previous publications. Here we discuss the
physical content of our results, and stress the generality of our
ideas.

{}From the physical point of view, our picture has three degrees of
freedom which fluctuate without an energy barrier. The spin flips are
generically massless, since spin up and spin down have the same energy
due to the global spin rotation invariance. The physical reason why
the transition between states with different local charges is
massless is more subtle, since it is not protected by any symmetry. In
fact the breaking of local U(1) gauge invariance is characteristic of
metallic state. The essential point\cite{skg} is that in order for an
{\it incoherent state} to be metallic, it is necessary to allow for
charge transfer between the localized degrees of freedom and the bath.
This can only happen if the local degree of freedom is in equilibrium
with the conduction electron bath. This requires the heavy level to be
at the chemical potential causing  the empty and the occupied states
to have the same energy.

This is reminiscent of recent approaches based on the slave boson
technique\cite{zou}. The slave boson describes the local charge
fluctuation and remains incoherent in the uncondensed phase. In both
treatments we see that, in the IMS, the local gauge invariance which
is broken by the coherent hopping is restored at low energies.

We  therefore find a unusual situation, that the impurity model that
the large d system maps onto  has a larger symmetry than one   would
have  naively expected. The metallic strongly correlated state is
locally described by an impurity model, a generalized Anderson model,
{\it in the mixed valence regime}.\cite{gk,sk}

\subsection{The Universal Low Energy Behavior}

{\bf Fixed Point Hamiltonian}
The generalized Anderson {\it impurity}  Hamiltonian
describes three local states coupled to an electron bath with a smooth
density of states:

\begin{eqnarray}
H= &&\sum_{k\sigma} (\epsilon_k - \mu) c_{k\sigma}^+c_{k\sigma}
+E^o_d d_{\sigma}^+d_{\sigma}
+{U \over 2} \sum_{\sigma \ne \sigma'} d^{+}_{\sigma} d_{\sigma }
d^{+}_{\sigma ' } d_{\sigma ' }\nonumber\\
&&+\sum_{\sigma} t ( d^{+}_{\sigma} c_{\sigma} + h.c. )
+ {V_1 } \sum_{\sigma,\sigma ' } d^{+}_{\sigma} d_{\sigma}
c^{+}_{\sigma ' } c_{\sigma '}
+ {V_2 \over 4} \sum_{\sigma_1,\sigma_2,\sigma_3,\sigma_4}
{\bf \tau}_{\sigma_1\sigma_2} \cdot {\bf \tau}_{\sigma_3\sigma_4}
d^{+}_{\sigma_1} d_{\sigma_2} c^{+}_{\sigma_3 } c_{\sigma_4}
\label{hamil.us}
\end{eqnarray}
where ${\bf \tau}$ label the Pauli matrices.
The local spin singlet $| s >$ and spin doublet $|\sigma>$ are
denoted by the vacuum $|0>$ with $E_0=0$ and the
singly occupied states $d_{\sigma}^+|0>$ with $E_{\sigma}=E_d^o$
respectively. An infinite onsite interaction, $U=\infty$, is
introduced to enforce the three dimensional {\it restricted}
configuration space. The hybridization $t$, the density-density
interaction $V_1$, and the spin exchange interaction $V_2$ describe
the generic couplings between the local states and the electron bath.

Here $\epsilon_k$ describes the dispersion of the electrons
in the bath. Such a dispersion is determined by solving the self
consistency condition. We envision that below a scale $t_p$  (the low
energy regime) the density of states is featureless and can be taken
to be a constant. We have shown that  such a non-singular density of
states is self-consistent.\cite{new}

At energy scales lower than $t_p$, we write the partition function and
the local correlation functions as Feynman sums over trajectories in
the local configuration space. We envision calculating N point
correlation functions of Hubbard operators $X_{\alpha, \beta}$ at N
values of imaginary time. The insertion of Hubbard operators force the
system to be at certain
configuration or to flip configurations at given values of imaginary
time. The amplitude for this process is the sum over all trajectories
consistent with these constraints. The weight for each trajectory
depends of course on the reaction of the electrons in the neighboring
sites to a given local trajectory.
This exact representation is due to Haldane.\cite{haldane}
It was recently extended by us, to incorporate generic local
configurations and couplings in all channels, using a simple
bosonization procedure. It has the form:

\begin{eqnarray}
{Z \over Z_0}
= \sum_{n=0}^{\infty} \sum_{\alpha_1,...,\alpha_n}
exp(-S[\tau_1, ... \tau_n ] )
\label{sumoverhis}
\end{eqnarray}
where

\begin{eqnarray}
S[\tau_1, ... \tau_n ] = &&\sum_{i<j}
(K(\alpha_i, \alpha_j) + K(\alpha_{i+1}, \alpha_{j+1})
- K(\alpha_i, \alpha_{j+1}) - K(\alpha_{i+1}, \alpha_{j}) )
ln {(\tau_j - \tau_i) \over \xi_o}\nonumber\\
&&- \sum_i ln (y_{\alpha_i\alpha_{i+1}})
+\sum_{i}h_{\alpha_{i+1}} {(\tau_{i+1}-\tau_i) \over \xi_o}
\label{hisaction}
\end{eqnarray}
Here $\tau_i$, for $i=1,...,n$, labels the hopping event from local
state $|\alpha_i>$ to $|\alpha_{i+1}>$, and $\xi_o$ is the ultraviolet
inverse energy cutoff.

The fugacities $y_{\alpha,\beta}$ describe the
amplitudes for quantum hopping between the configurations
$|\alpha>$ and $|\beta>$. The charge fugacity describes the
hopping between two local states with different charges and is
determined by the hybridization term, $y_{0,\sigma}=y_t=t\xi$.
Similarly, the spin fugacity describes the hopping between two local
states with different spin quantum numbers and is determined by the
transverse component of the exchange coupling, $y_{\sigma \ne \sigma
'} = y_j = {V_2^{\perp} \over 2 } \xi$.
The fields $h_{\alpha}$ describe the energy splitting among the local
configurations. They have to be introduced since generically
there is no symmetry between the local many body configurations. In
our three state problem, there exists a spin rotational symmetry but
not a symmetry between spin and charge. Therefore, we introduce a
field related to the $d$ level $E_d$, i.e. $h_0= - {2 \over 3} E_d\xi$
and $h_{\sigma}= {1 \over 3} E_d\xi$.

The logarithmic interaction between the hopping events (which can be
thought of as defects) describe the reaction of the electron bath
towards disturbances at the impurity sites. The bare values of the
stiffness constants, $\epsilon_t=-K(0,\sigma)$ and
$\epsilon_j=-K(\sigma,\sigma' \ne \sigma )$ are determined by the
interaction strengths. They provide the initial conditions of the
renormalization group flow, and will be discussed in detail in the
next section. The logarithmic interaction in the action
(\ref{hisaction}) can be written as a classical ``spin'' model with
long range ${1 \over {\tau}^2}$ interaction, $\sum_{i<j}
K(\alpha_i,\alpha_j) ({\xi_0 \over \tau_i-\tau_j})^2$.\cite{cardy}
This allows us to adapt Cardy's RG equations\cite{cardy} to our
problem.

{\bf Scaling Equations} The renormalization group equations describe
how the fugacities evolve as a function of length scale.  A detailed
derivation can be found in Refs. \cite{sk}. Here we quote the results
for the spin ${1 \over 2}$ case,

\begin{eqnarray}
&&d y_t /d ln \xi =
(1- \epsilon_t)y_t + y_t y_j\nonumber\\
&&d y_j / d ln \xi =
(1- \epsilon_j)y_j + y_t^2\nonumber\\
&&d \epsilon_t / d ln \xi =
-6\epsilon_ty_t^2
+\epsilon_j (y_t^2 - y_j^2)\nonumber\\
&&d \epsilon_j / d ln \xi
= -2\epsilon_j (y_t^2+2 y_j^2)\nonumber\\
&& d {E_d\xi} / d ln \xi = (y_t^2 - y_j^2 )
+E_d \xi (1-3 y_t^2)
\label{scaling}
\end{eqnarray}

In the renormalization of the fugacities, the linear terms give the
associated {\it anomalous dimensions}, while the quadratic terms
reflect the non-abelian nature of our three state problem. The
renormalization of the stiffness constants reflect the correction
to interactions induced by the fugacities. Finally, in the
renormalization of the energy level, the $y_t^2$ and $y_j^2$ terms
arise due to the particle-hole asymmetry.\cite{haldane}

{\bf Universality Classes}
The defects are bound together when the attraction among
them are strong enough. This leads to fugacities which are
renormalized to zero. As in the Kosterlitz-Thouless transition in the
XY magnet, unbinding of the defects occurs when the attraction becomes
weaker. In our case, $\epsilon_t$ and $\epsilon_j$ reflect the
strength of the long range interactions between the defects.

Since the spin and charge defects are coupled, the unbinding
of these defects can occur at the same time, or at different stages.
We find that, there can be three kinds of fixed points.

{\it The strong coupling mixed valence fixed point} occurs when both
the spin and charge defects are unbound. Here both the hybridization
and the Kondo exchange are relevant, and therefore both the local spin
and local charge degrees of freedom are quenched. This state is in the
same universality class as the usual strong coupling phase of the
Anderson model. The basin of attraction of this case is given by
$\epsilon_t < 1$ and a range of $\epsilon_j < 1$ when $\epsilon_t >
1$.

{\it The weak coupling mixed valence fixed points} occur when both the
spin and charge defects are bound. Here both the hybridization and the
Kondo exchange are irrelevant, and therefore neither the local spin
nor the local charge degrees of freedom are quenched. The basin of
attraction for this case is determined by $\epsilon_t > 1$ and
$\epsilon_j > 1$.

{\it The intermediate coupling mixed valence fixed points} occur when
the spin defects are unbound while the charge defects are
bound.\cite{note2} The hybridization is irrelevant, while the Kondo
exchange is relevant. Therefore, local spins are quenched while local
charges are not. The basin of attraction of this case is in a range
within $\epsilon_t > 1$ and $\epsilon_j < 1$.  Such an intermediate
phase is the analog of the hexatic phase in the two dimensional
melting problem, in which unbinding occurs for dislocations but not
for disclinations.\cite{melting}

As the chemical potential is varied, the strong coupling mixed valence
state occurs within a crossover region over a range of chemical
potential. This range is determined by the renormalized hybridization,
as was characterized by Haldane.\cite{haldane} For the weak coupling and
intermediate coupling cases, the renormalized  hybridization is zero.
Therefore, the mixed valence states occur at a critical chemical
potential $\mu_c$. As $\mu$ deviates from $\mu_c$, the mixed valence
states eventually crossover into either an empty orbital or a local
moment regime. It should be emphasized that, the pinning effect
discussed previously led to the conclusion that, $\mu_c$ corresponds
to {\it a range of total electron densities}.

Within the strong coupling mixed valence state, an energy scale
(renormalized Fermi energy) is generated below which quasiparticles
appear. The low energy physics is described by a coherent Fermi
liquid.

In the weak coupling mixed valence state, no such energy scale is
generated, and the system is scale invariant. The renormalized Fermi
energy vanishes, resulting in a non-Fermi liquid state--an incoherent
metallic state. Local correlation functions, both in spin and charge
channel, show algebraic behavior. Spin and charge excitations are
asymptotically decoupled.

In the intermediate coupling mixed valence state, a coherent energy
scale is generated in the spin channel, but not in the charge
channel. We have therefore a spin-charge separated state with coherent
spin excitations and incoherent charge excitations. This state
resembles the spin-charge separated state that occurs in the gauge
approach when slave boson is not
condensed,\cite{zou}
which has coherent `spinon' excitations and incoherent
`holon' excitations.

{\bf Critical Behavior}
The description of the physics in terms of defects allows us to make
several general statements about the {\it zero temperature quantum
phase transitions} between the various states we defined. These are
binding unbiding transitions in $0+1$ dimensions.

The transitions are characterized by the collapse of an energy scale,

\begin{eqnarray}
\epsilon_F \sim \epsilon_0 {\rm e}^{-{1 \over (\epsilon -
\epsilon_c)^{\eta}}}
\label{critical}
\end{eqnarray}

For the transition between the Fermi liquid and the weak coupling
non-Fermi liquid states, the binding-unbinding transition for spin
defects is driven by that of the charge defects. Therefore, $\epsilon_F$
corresponds to the Fermi energy of the Fermi liquid.
For the two stage transition from the Fermi liquid, through the
intermediate coupling non-Fermi liquid state, to the weak coupling
non-Fermi liquid state, the $\epsilon_F$ corresponds to the coherence
energies in the spin channel and in the charge channel respectively.
The exponent $\eta={1 \over 2}$ for the transitions from the weak
coupling phase to the strong coupling phase and from the intermediate
coupling phase to the strong coupling phase, while $\eta=1$ for the
transition from the weak coupling phase to the intermediate coupling
phase due to rotational invariance.

\subsection{The Non-Universal High Energy Behavior--Quantum Chemistry}

The low energy behavior of a metallic system is universal in the sense
that the behavior of the local correlation functions is controlled by
the different fixed points of the impurity hamiltonian described
in the previous section.

In a previous publication we arrived at that impurity model starting
 from an extended two band Hubbard model on a lattice in the limit of
infinite dimensions \cite{sk}. We stress however that the {\it low energy}
description of the previous section is in fact much more general.
There are many lattice hamiltonians, whose {\it local physics at low
energies} is described by the renormalization group equations
(\ref{scaling}). In particular  one can arrive at low energies at  the
impurity model of the previous section starting from a metallic {\it
one band} system.  The role of the  ``light electrons'' of Ref.
\cite{sk} is now played by the incoherent part of the one particle
Green's function of the one band system.

In fact one can arrive at our {\it local low energy} description for a
correlated  {\it metallic state} from very general considerations. The
local configurations are expected to have different spin and charge
quantum numbers, and can be described by a spin doublet and a spin
singlet. The electron bath should have non-zero density of states
near the Fermi level. They are coupled generically in both spin and
charge sectors. Finally we also expect a one to one correspondence
between the local degrees of freedom and the degrees of freedom of
the bath, as embodied in the self consistency condition.

While these assumptions are simple and natural they are
by no means exclusive. Ruckenstein, Varma and collaborators have
suggested  that an impurity model with a bath with {\it several
channels} of screening electrons is related to the copper
oxides.\cite{giamarchi} These impurity models  have not been {\it
derived} from a lattice model in large dimensions.

We now turn to the crucial question of the determination of the
initial conditions for our RG flows. This is a non universal high
energy question that depends on system to system.

In an extended Hubbard model with the Lorentzian case, the width of
the scaling regime $t_p$ equals the bare conduction electron
bandwidth. The initial conditions are determined by the phases
shifts of the conduction electron bath, which are directly related to
the physical interaction strength. They are given as follows,

\begin{eqnarray}
&&\epsilon_t={1 \over 2}[ ( 1 - {\delta_2 \over {\pi}}
- {\delta_1 \over {\pi}})^2+ ( {\delta_1 \over {\pi}}-{\delta_2 \over
{\pi}})^2 ]\nonumber\\
&&\epsilon_j= ( 1 - 2 {\delta_2 \over {\pi} } )^2
\label{stiffness}
\end{eqnarray}
where the phase shifts are $\delta_1=tan^{-1}(\pi \rho_oV_1)$ and
$\delta_2=tan^{-1}(\pi \rho_o V_2/4)$, with $\rho_o$ the conduction
electron density of states at the Fermi level.

The basin of attraction for different phases given in the previous
section can now be specified in terms of the phase shifts. The weak
coupling non-Fermi liquid state occurs $\delta_1 / \pi < -(\sqrt{3} -1
)/2$ and  $\delta_2 < 0$. The intermediate coupling non-Fermi liquid
state occurs over a range within $\delta_1 / \pi < -(\sqrt{3} -1 )/2$
and  $\delta_2 > 0$. Otherwise, a Fermi liquid state emerges at low
energies.

To determine the basin of attraction of the various phases
in more realistic models, a more elaborate treatment of the high
energy region is necessary. One has to solve first the self
consistency equation for the parameters of the impurity model, and
then integrate out the high energy states of the impurity to arrive at
the low energy effective impurity model. Previous work on the Hubbard
model demonstrated that, the impurity models that describe the local
physics of lattice problems are characterized by atomic like features
(Hubbard bands) and narrow resonances near the Fermi level. To treat
the atomic like features  in a narrow band situation we can carry out
a poor man's version of a quantum chemistry calculation to determine
the initial conditions for the couplings that enter the low energy
renormalization group flow.

To illustrate our idea, we parametrize the associated impurity
problem, i.e. the $\Gamma_{\alpha}$ in $S_{eff}$ using the following
Hamiltonian:

\begin{eqnarray}
{\rm H_{imp}}=&& \epsilon_d d_{\sigma}^+d_{\sigma} + {U \over 2}
\sum_{\sigma \ne \sigma'} d^{+}_{\sigma} d_{\sigma } d^{+}_{\sigma ' }
d_{\sigma ' } +\epsilon_p p_{o,\sigma}^+p_{o,\sigma}  +
\sum_{<i,j>} t^p_{ij} (p_{i \sigma}^+ p_{j \sigma} + h.c. )\nonumber\\
&& - t (d_{\sigma}^+ p_{o \sigma} + h.c. )
+V_1^o \sum_{\sigma , \sigma '} d_{\sigma}^+  d_{\sigma'} p_{\sigma '}^+
p_{\sigma '} + V_2^o \sum_{\sigma , \sigma '} d_{\sigma}^+ d_{\sigma '}
p_{\sigma '}^+ p_{\sigma }
\label{hamil}
\end{eqnarray}
Here $\epsilon_d$ and $\epsilon_p$ label the levels of the local $d$
and $p$ electrons, both of which are determined by the corresponding
levels and the chemical potential of the lattice model.
The Hubbard interaction $U$ is taken as infinite. $t^p_{01}$ reflects
the coupling between $p$ electrons at the local and the nearest
neighbor site. $t^p_{ij}$ parametrize the generic spectral function of
the electron bath. We have also included the generic interaction
parameters between the $d$ and $p$ electrons, a {\it repulsive}
interaction $V_1^o$ in the density-density channel, and an {\it
antiferromagnetic} interaction $V_2^o$ in the exchange channel.
For a {\it bounded} density of states, and when the hybridization
is large we expect the inequality $t > t_p$.

The scaling regime occurs within the energy range of $t_p$ near the
Fermi level of the electron bath.
At energy scales larger than this bandwidth $t_p$, we cannot apply the
logarithmic RG based on decimation of the conduction electron
bandwidth and perform instead a poor man's version of a realistic
quantum chemical calculation. We first ignore $t_p$ and diagonalize
the local problem and label the twelve  eigenstates  by the charge $N$
and the spin quantum numbers. The eigenvectors and the eigenvalues are:

\begin{quasitable}
\begin{tabular}{cccc}
& &&\\
\FL{$|N=0>$}&$\epsilon (N=0) = 0$&\\
$|N=1, \sigma, b> = ( u_1 d_{\sigma}^+ + v_1 p_{\sigma}^+)|0>$&
$E (N=1, b) = {1\over 2} (\epsilon_p + \epsilon_d - R_1)$\\
$|N=1, \sigma, ab> = ( -v_1 d_{\sigma}^+ + u_1 p_{\sigma}^+)|0>$&
$E (N=1, ab) = {1\over 2} (\epsilon_p + \epsilon_d + R_1)$\\
$|N=2, s,b> = ( u_2 \epsilon_{\sigma \sigma '}{1 \over \sqrt{2}}
d_{\sigma}^+ p_{\sigma'}^+ + v_2 p_{\uparrow}^+ p_{\downarrow}^+ )
|0>$ & $E (N=2, s, b) = \epsilon_p
+{1\over 2} (\epsilon_p + \epsilon_d +V_1^o -V_2^o- R_2)$\\
$|N=2, s,ab> = ( -v_2^o \epsilon_{\sigma \sigma '}{1 \over \sqrt{2}}
d_{\sigma}^+ p_{\sigma'}^+ + u_2 p_{\uparrow}^+ p_{\downarrow}^+ )
|0>$ & $E (N=2, s, ab) = \epsilon_p +{1\over 2} (\epsilon_p +
\epsilon_d +V_1^o -V_2^o + R_2)$\\ $|N=2, t> = d_{\sigma}^+
p_{\sigma}^+ |0>,~ ~ ~ {1 \over \sqrt{2}} \sum_{\sigma} d_{\sigma}^+
p_{- \sigma}^+ |0>$ & $E (N=2, t > = \epsilon_p + \epsilon_d +
V_1^o+V_2^o$\\
$|N=3, \sigma> = d_{\sigma}^+ p_{\uparrow}^+ p_{\downarrow}^+ |0>$ &
$E (N=3) = \epsilon_d +2 \epsilon_p + 2 V_1^o$
\end{tabular}
\end{quasitable}
\begin{eqnarray}
\label{states}
\end{eqnarray}
Here ``b'',``ab'', ``s'' and ``t'' refer to bonding and antibonding
combinations and singlet and triplet states respectively.
The coherence factors are $u_1^2=1-v_1^2={1\over
2}(1+\Delta/R_1)$ and $u_2^2=1-v_2^2={1\over
2}(1+(\Delta-V_1^o+V_2^o)/R_2)$, where $\epsilon_p-\epsilon_d=\Delta$,
$R_1 = \sqrt{\Delta^2 + 4 t^2}$, and $R_2 = \sqrt{(\Delta-V_1^o+V_2^o)^2 +
8 t^2}$.

As we vary the chemical potential, the energies of these
local states vary. We consider the hole-doped regime,
in which case $\epsilon_p$ is such that the local doublet $|N=1, \sigma, b>$
and the local Zhang-Rice-like singlet $|N=2, s,b>$ are almost degenerate,
$E (N=1, b) \approx E (N=2, s,b)$.
Such a mixed-valence condition determines $\epsilon_p$ to be
$\epsilon_p \approx {1 \over 2}(R_2 - R_1 + V_2^o - V_1^o)$,
which in turn specifies the energies for various excited states
$\gamma$: $\epsilon (\gamma)= E (\gamma) -  E (N=1, b)$. They
are given as follows,

\begin{eqnarray}
&&\epsilon  (N=0) = {1\over 2} (2R_1+\Delta -R_2 + V_1^o
-V_2^o)\nonumber\\
&&\epsilon  (N=1, ab) = R_1\nonumber\\
&&\epsilon  (N=2, s, ab) = R_2\nonumber\\
&&\epsilon  (N=2, t ) = {1 \over 2} (R_2 -\Delta + V_1^o + 3 V_2^o
)\nonumber\\
&&\epsilon  (N=3) = {1\over 2} (2R_2-\Delta -R_1 + 2 V_1^o +2 V_2^o)
\label{exc.energy}
\end{eqnarray}

We now integrate out approximately the high energy eigenstates using
second order perturbation theory to construct the low energy
effective Hamiltonian:

\begin{eqnarray}
{\rm H_{eff}} = &&\epsilon_d (\sum_{\sigma} |\sigma><\sigma| - |s><s|)
+\tilde{t} (\sum_{\sigma} |\sigma> <s | p_{-\sigma}^+ sgn(\sigma) +
h.c.)\nonumber\\
&&-\tilde{V}_1 \sum_{\sigma , \sigma '} |\sigma><\sigma | p_{\sigma
'}^+ p_{\sigma '} + {\tilde{V}_2 \over 4}
\sum_{\sigma_1,\sigma_2,\sigma_3,\sigma_4}
{\bf \tau}_{\sigma_1\sigma_2} \cdot {\bf \tau}_{\sigma_3\sigma_4}
|\sigma_1><\sigma_2 | p_{\sigma_3}^+ p_{\sigma_4 }
\label{effham2}
\end{eqnarray}
with the various  parameters  given by:
\begin{eqnarray}
&&\tilde{V}_1 = {1 \over 2}(t^p_{01})^2 (-{v_1^2 \over \epsilon (N=0)}
+ {3 \over 2} {u_1^2 \over \epsilon (N=2,t)} - {u_2^2 \over \epsilon
(N=3)} + {2 \over \epsilon (N=1,ab)} (v_2u_1 - { v_1 u_2 \over
\sqrt{2}})^2 )\nonumber\\
&&\tilde{V_2} = {1 \over 2}(t^p_{01})^2 ({v_1^2 \over \epsilon (N=0)}
- {1 \over 2} {u_1^2 \over \epsilon (N=2,t)} + {1 \over \epsilon (N=2,
s, ab)} (v_1u_2 - { u_1 v_2 \over \sqrt{2}})^2)\nonumber\\
&& \tilde{t} = v_1 t
\label{parameter}
\end{eqnarray}

By using a suggestive notation, we define a new vacuum
$|vac>=s^+|0>$, and creation operators for the doublet $d_{\sigma}^+ =
|\sigma><s|$ which obey a no double occupancy constraint. After
performing a particle-hole transformation $p_{-\sigma}^+ sgn(\sigma)
\rightarrow p_{\sigma}$, the effective Hamiltonian given in Eq.
(\ref{effham2}) reduces to the  Hamiltonian, Eq.
(\ref{hamil.us}) described in the previous
section   with $V_1=\tilde{V_1}$ and $V_2=\tilde{V_2}$.
We can now apply the RG analysis to determine the basin of attraction of
the various phases.
As stated above it is natural to express the answers  in terms of the
effective phase shifts $\tilde{\delta_1}=tan^{-1}(\pi \rho \tilde{V_1})$
and $\tilde{\delta_2}=tan^{-1}(\pi \rho \tilde{V_2})$.
The system renormalizes into the weak coupling non-Fermi liquid state
if $\tilde{\delta_1}/\pi < -(\sqrt{3} -1 )/2$ and
$\tilde{\delta_2} < 0$, and into the intermediate coupling non-Fermi
liquid state if $\tilde{\delta_1}/\pi < -(\sqrt{3} -1 )/2$ and
$\tilde{\delta_2} > 0$.

The effective interactions, given in Eq. (\ref{parameter}),
depend on the way high energy states arrange among themselves which in
turn is determined by the {\it quantum chemistry}
(i.e. the initial parameters in the full impurity
model, Eq. (\ref{hamil})) of the problem. The basin of attraction of
the non-Fermi liquid states occurs when $\epsilon (N=0)$ or $\epsilon
(N=3)$ is small. It is easy to see from Eq. (\ref{hamil}) that,
$\epsilon (N=0)$ is small for a strong antiferromagnetic $V_2^o$ and a
weak repulsive $V_1^o$. We explored the parameter space  $t$, $V_1^o$
and $V_2^o$ for a given $\Delta$, and  found that  for large $V_2^o$
and small $V_1^o$ the non-Fermi liquid regime is larger when the
hybridization increases.

In Fig. \ref{int}, we plot $\tilde{V_1}/(t_{01})^2$ and
$\tilde{V_2}/(t_{01})^2$ as a function  of $V_2^0$ for a $\Delta=2$,
$t=1$, and $V_1^0=0$. It can be seen that, $\tilde{V_1}/(t_{01})^2$
turns to negative when $V_2^0$ is large enough, and diverges upon
approaching $(V_2^0)_c \approx 2.415$ at which $\epsilon (N=0)=0$.
This divergence is not physical but indicates the breakdown of second
order perturbation theory. Qualitatively we can say that the
intermediate phase is realized when the quantum chemistry of the
problem is such as to have a low lying excited ``empty sate''. This
happens when the Kondo exchange which stabilizes the Zhang-Rice
singlet is large and when the hybridization $t$ is comparable to
$\Delta$.

Finally, we observe that, in finite dimensions, the incoherent
metallic states cannot be stable at zero temperature. The algebraic
decay of the local correlation functions are symptomatic of a
superconducting instability that sets in at a finite energy scale
$T_c$. The feedback of the long range order over the single particle
properties is down by powers of ${1 \over d}$. In finite dimensions
the coupling of the long range order on the local physics modifies the
local effective action at temperatures $T < T_c$, and has the effect
of cutting off the power law singularities. Nevertheless, the
incoherent metallic states have a well defined existence at
temperatures $T_c << T << \omega_c$ in the same sense that the
paramagnetic Mott insulating phase is well defined only for $T_N << T
<< \epsilon_F$, $T_N$ being the Neel temperature.

The ideas presented here approach the problem of the breakdown of
Fermi liquid in strongly correlated fermion systems
in the same spirit that the combination of the Anderson-Yuval RG
calculation coupled to the Nozieres strong
coupling fixed point analysis solves the antiferromagnetic Kondo
problem. It provides a conceptual framework, in which one solves a
complicated problem by treating it with different tools at different
energy scales. The Kondo problem involves the spontaneous generation
of a scale. Therefore it is self-similar at short distances, and
requires strong coupling methods at long distances. Here we find
ourselves in the opposite situation: the problem is to understand the
generation of a scale invariant self-similar low energy regime. The
strong coupling methods are applied at high energy scales, and the
renormalization group, when started with the correct quantum chemistry
initial conditions, drives us towards the weak coupling self-similar
regimes.

ACKNOWLEDGMENTS:
We thank J. Cardy, A. Finkelstein, A. Georges, T. Giamarchi, A.
Ruckenstein, and C. Varma for useful discussions. This work was
supported by the NSF under grant DMR 922-4000.

\newpage

\figure{Effective interactions $\tilde{V_1} /(t_{01}^p)^2$ (solid
line) and $\tilde{V_2} /(t_{01}^p)^2$ (dashed line) as a function of
the ``bare'' interaction $V_2^0$ defined in the text. The parameters
are $\Delta=2$, $t=1$, and $V_1^0=0$.\label{int}}


\begin{references}
\bibitem[16]{note2} The scaling equations show that, it is impossible
to have spin kinks bound while charge kinks are unbound.
\end{references}
\end{document}